# White Paper
# Brief overview of current practices for open consultation

Vassilis Giannakopoulos[1], George Giannakopoulos [1,2]
[1]SciFY PNPC and [2]NCSR "Demokritos"

A SciFY White Paper on eDemocracy

# 1. Introduction

The purpose of this document is to provide a brief overview of open consultation approaches in the current, international setting and propose a role for Information Technologies (IT) as a disruptive force in this setting. The document is structured as follows. We describe the motivation behind this work in Section 1.1; we identify trends and challenges in Section 1.2. Then, we look deeper into the first stages of public consultation in Section 2, referring to current practices across the globe. We conclude the text with a summary of the findings and proposed next steps in Section 3.

## 1.1 Motivation and scope

The public consultation process in Greece faces many issues: research showed that although people find it very valuable, trust towards the process is low and citizen participation has been declining. OGP commitments to strengthen the public consultations were clear, and progress has been made but technical progress in the implementation has been slow.

SciFY has worked towards facing the above challenges through e-democracy tools. In order to best understand the international scenery and adapt our approach, we conducted a study of current open consultation practices, which we share in this document, as part of our open knowledge dissemination action.

The results of this study empowered the creation of a innovative, even disruptive, tool: DemocracIT. DemocracIT, is an innovative public consultations platform that allows policymakers engage with citizens at the final stage of the consultation process in an effective way. It provides a rich annotation and discussion environment, coupled with powerful reporting mechanisms exploiting data mining methods. It includes cross-consultation statistics and analysis to quantify the way organizations follow (or do not follow) best practices regarding open consultation. The platform is developed as an open source project to maximize reuse and sustainability of the project.

With DemocracIT we respond to the need for more transparency, higher citizen participation and clearer impact measurement. Its open source approach maximizes reuse potential and, thus, may empower communities beyond the traditional law-making ones, including possibly cross-country policy making organizations and settings. The open source approach allows for adaptability, customization, and therefore international use.

## 1.2 Trends and Challenges

Citizen engagement in the decision-making process has significantly increased during the last two decades **[1]**. A number of reasons can explain this trend: the complexity of the issues to be solved, the knowledge citizens and citizen groups possess around a problem, the need for acceptance of the policies to be implemented. But it is emerging ICT technologies, Internet usage and web 2.0 tools that are the technical enablers, since they have made wider citizen participation easier. Yet, our research around practices and tools used for public consultation, a function that permeates all the stages of the policy making cycle, shows that a number of challenges still need to be addressed:

**– Inadequate use of ICT technologies.** Public consultation practices require that policymakers deal with large amounts of textual input. Although ICT technologies (e.g. text mining, sentiment analysis) exist, they are not exploited to allow for an effective analysis of the available citizen feedback, making the work of the policymakers more difficult.

**– Public consultation of the final draft laws is a real need, yet the approach is inadequate:** at this stage, the only way citizens can leave their comments is through e-mails and forms. This approach is ineffective in many ways: it does not allow discussion, interaction and mutual understanding, nor does it encourage participation.

**– Lack of an evaluation mechanism for the consultation process:** There is no way for the citizens to check what their contributions were, and if and how they got into consideration or incorporated in the final text of the law. This increases the lack of trust of the public on the outcome of e-participation processes **[2, 3]**.

# 2. Public Consultation at the first stages of the policymaking process

E-democracy is divided into three sub-fields: "information provision, deliberation, and participation in decision-making" **[1]**. Civic engagement includes three dimensions: political knowledge of public affairs, political trust for the political system, and political participation in influencing the government and the decision-making process **[4]**. The Internet provides a new avenue to interact with governmental institutions **[5]**. In the above context, different levels of government (e.g. municipalities, states, state unions) in different continents have included ways to allow citizens and/or other players in the field (e.g. political organizations, trade associations etc.) to get involved in the process of policymaking.

The policymaking process includes different steps that could be broadly described as follows:
– Issue identification / Agenda setting
– Policy formulation
– Decision making
– Implementation
– Evaluation

Yet, the above stages are indicative; the lawmaking process is usually iterative and more systemic. Consultation can permeate the entire process, to maximally integrate feedback and improve the resulting policies.

**We looked into practices of different governmental structures regarding the first three steps of the policymaking process.** For these different stages, different e-participation tools are being used.

## 2.1 Stage 1: Issue identification / Agenda setting

This stage includes problem identification and quantification. It includes the following actions to allow participation:

| Action | Tools | Examples |
| --- | --- | --- |
| Problem identification / quantification | ● Blogging-like tools,<br>● "Post your story",<br>● Discussion forums | 1. Australian Government<br>2. UK government |

## 2.2 Stage 2: Policy formulation

This stage usually includes the following actions to allow participation:

| Action | Common Tools | Examples |
| --- | --- | --- |
| Collecting ideas | Various tools that allow citizens to submit ideas and vote on them | Australian Government |
| Getting feedback on different suggested policies / priorities | Structured questionnaires | 1. European Union<br>2. Victoria, Australia<br>3. Canada |
| Budget allocation | Specific web tools | YouChoose (UK) |

## 2.3 Stage 3: Decision making

In this stage, citizens usually participate by giving feedback on the final draft law.

| Action | Tools | Examples |
| --- | --- | --- |
| Collecting feedback on the final text of the draft law | e-mails | 1. Business consultations portal in Australia<br>2. UK Public Consultation site |
| | Forum-like tools that allow for remarks | Greek OpenGov portal for open consultations |
| | Specific forms with general remarks (with personal info needed) | USA public consultation portal |

There is a variety of tools and methods to gather feedback for the first stages of the lawmaking process. For example, the NOMAD platform **[6]** allows a policy maker to gather feedback from the Web and Social Media to improve a policy proposal. The PADGETS system **[7]** supports policy makers by combining social media analysis and simulation to provide feedback on policy making. However, as indicated above the use of technology is very limited in the public consultation of the final text versions of the draft law. Furthermore, in most cases the tools function as collectors of data and do not always empower the citizen; they empower the policy maker.

Yet, the need for a system that enables citizens and lawmakers to interact efficiently during the last stages of the lawmaking process is clear. For example, the Australian Government states that automated tools and techniques for evaluating online submissions are much needed, because some consultations receive a large volume of submissions. For example, the 2009 National Human Rights Consultation , which included an online element, received over 35,000 submissions. It suggests the "use of text mining tools" and the "design the online submission mechanisms in such a way that there is already some level of meaning attached to content at the point of submission, for example by asking contributors to tag or categorise their submissions based on the topic/s of the consultation". The EU does not have a tool to address this stage of the consultation.

## Other consultation platforms

There exist tools used for choosing priorities, setting budgets, brainstorming ideas and voting on them. Such solutions:
1. Allow commenting of small texts (ideas / challenges etc.)

2. Work like forums, where participants can comment, and/or agree with other people's ideas
3. Integrate Social Media into polling (Answer questions and post answers to Social Media)

Probably the most representative ones are:

[CitizenSpace](): It allows for online surveys, uploading of files from respondents, manually entering answers received outside the public consultations (e.g. through mail or email)

[EngagementHQ](): Forum-like tool

It has to be noted though that these tools are not built to allow people to discuss on specific texts, at a later/final step of a process, where the final text of the draft law is open for consultation.

## 3. Conclusions

We consider that Information technology tools can significantly empower democratic action and collaboration, but, based on the above study, they need to:
- remain user-oriented,
- be well-designed and efficient,
- be taking into account all the stakeholders and
- take advantage of advances in ICT.

In ancient Greece, "[Ecclesia]()" had the final say on legislation, through direct participation of the citizens. Now, the challenges of space and time, the number and motives of participants have to be well-addressed. The new "[Demos]()" needs new tools to function in an inclusive and effective way, in order to strengthen Democracy and participation.

DemocracIT provides new tools that address these issues. Being open source and modular, it allows for implementation in different countries and settings. It can be combined with other tools to allow for the creation of new solutions that serve the people.

## Acknowledgements

DemocracIT was funded by Iceland, Lichtenstein and Norway in terms of the program [«We Are All Citizens»](), which is part of the Funding Mechanism of EEA for Greece, also known as [EEA Grants](). Administrator of the subvention of the Program is [Bodossaki Foundation](). Aim of the Program is the empowering of the society of the citizens in our country and the aid of social justice, democracy and sustainable development.

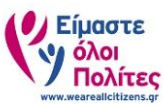
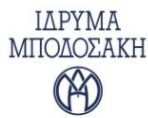
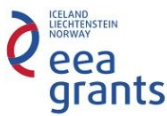

---